# Current-Induced Magnetization Switching by the High Spin Hall Conductivity α-W

*Wei-Bang Liao, Tian-Yue Chen, Yari Ferrante, Stuart S. P. Parkin, and Chi-Feng Pai\**

Wei-Bang Liao, Tian-Yue Chen, Prof. Chi-Feng Pai
Department of Materials Science and Engineering
National Taiwan University
Taipei 10617, Taiwan
E-mail: cfpai@ntu.edu.tw

Dr. Yari Ferrante
IBM Research-Almaden
San Jose, CA, 95120, USA

Prof. Stuart S. P. Parkin
Max Plank Institute for Microstructure Physics
Weinberg 2, 06120, Halle (Saale), Germany

The spin Hall effect originating from 5d heavy transition metal thin films such as Pt, Ta, and W is able to generate efficient spin-orbit torques that can switch adjacent magnetic layers. This mechanism can serve as an alternative to conventional spin-transfer torque for controlling next-generation magnetic memories. Among all 5d transition metals, W in its resistive amorphous phase typically shows the largest spin-orbit torque efficiency ~ 0.20-0.50. In contrast, its conductive and crystalline α phase possesses a significantly smaller efficiency ~ 0.03 and no spin-orbit torque switching has yet been realized using α-W thin films as the spin Hall source. In this communication, through a comprehensive study of high quality W/CoFeB/MgO and the reversed MgO/CoFeB/W magnetic heterostructures, we show that although amorphous-W has a greater spin-orbit torque efficiency, the spin Hall conductivity of α-W ( $\left|\sigma_{SH}^{\alpha\text{-}W}\right| = 3.71\times10^5 \ \Omega^{-1}\text{m}^{-1}$ ) is ~3.5 times larger than that of amorphous-W ( $\left|\sigma_{SH}^{\text{amorphous-}W}\right| = 1.05\times10^5 \ \Omega^{-1}\text{m}^{-1}$ ). Moreover, we demonstrate spin-orbit torque driven magnetization switching using a MgO/CoFeB/α-W heterostructure. Our findings suggest that the conductive and high spin Hall conductivity α-W can be a potential candidate for future low power consumption spin-orbit torque memory applications.

When passing an electric current through heavy transition metals (HM) with strong spin-orbit coupling, a transverse pure spin current can be produced due to spin-dependent scattering events: this phenomenon is called the spin Hall effect (SHE)[1-4]. By placing a ferromagnetic layer (FM) in contact with a HM layer, the pure spin current originating from the SHE can be transmitted from the HM layer into the FM layer. The transmitted spin angular momentum can then be transferred to the local magnetic moments in the FM layer and generate spin-orbit torque (SOT) driven dynamics such as magnetization switching,[5, 6] magnetic oscillation,[7, 8] or domain wall motion.[9, 10] It has been previously shown that among pure 5d HMs, amorphous-W (or also β-W) has the largest damping-like SOT efficiency or spin Hall ratio of ~ 0.20-0.50.[11-14] However, while compared to other large SOT efficiency materials such as Pt-based alloys[15, 16] or conductive topological insulators,[17] amorphous-W is less favorable for applications due to its high resistivity ($\rho^{amorphous-W}$ ~ 200-300 μΩ-cm) and, therefore, greater power consumption to achieve current-induced SOT-driven magnetization switching. Interestingly, the possibility of employing conductive and crystalline α-W (having about an order of magnitude smaller resistivity, $\rho^{\alpha-W}$ ~ 20-40 μΩ-cm) for SOT applications has yet to be explored, presumably due to its small reported SOT efficiency of ~ 0.03.[18] Nevertheless, this magnitude of efficiency is not negligible and SOT switching in an α-W/FM heterostructure should be achievable with suitable thin film stacks design and optimization. More importantly, if the SOT switching in such heterostructures can be achieved, the highly-conductive nature of crystalline α-W makes it a potential candidate for low-power-consumption spintronic device applications.

To study the SOTs in W-based heterostructures systematically, we prepare two series of samples on amorphous Si/SiO₂ substrates using an ultra-high vacuum magnetron sputtering deposition chamber: a normal structure W($t_w$)/Co$_{20}$Fe$_{60}$B$_{20}$(1)/MgO(2)/Ta(2) and a reversed structure Ta(5)/Co$_{20}$Fe$_{60}$B$_{20}$(0.3)/MgO(2)/Co$_{20}$Fe$_{60}$B$_{20}$(1.2)/W($t_w$)/Ta(3) (numbers in the parenthesis are in nanometers) with $t_w$ varying (by 1nm) from 2 to 10 nm. Both series of samples are post-annealed at 300 °C for 1 hour in a high-vacuum anneal chamber in order to promote perpendicular magnetic anisotropy (PMA) of the CoFeB layers. Note that, in the reversed structure, the amorphous and non-magnetic Ta(5)/CoFeB(0.3) bilayer is used as a smooth oxygen-free underlayer before the growth of MgO.

It is known that thin W ($t_w \leq 6$ nm) layers used in the normal structure are typically amorphous while thicker W ($t_w > 6$ nm) layers are usually crystalline (α-W)[11, 14]. To confirm whether the structure of W in our samples is consistent with previous works and that the W films in our reversed structures show a similar structural phase transition with increasing thickness, out-of-plane θ-2θ X-ray diffraction (XRD) measurements are first performed (**Figure 1**a and b). It is found that in the normal structure the thin W layers ($t_w \leq 4$ nm) are likely amorphous while the thicker ($t_w > 4$nm) W layers possess a crystalline α phase with a (110) texture, which is fairly consistent with the trends previously reported. However, in the reversed structure, the α-phase diffraction peaks can be observed for $t_w = 4$ nm, which is in the relatively thin regime compared to the normal structure case. To verify these findings we perform transmission electron microscope (TEM) imaging of both structures with $t_w = 4$ nm. The cross-sectional TEM results show that all the deposited layers are uniform (Figure 1c and d). Also, by fast fourier transform (FFT) analysis, the W(4) buffer layer is found to be amorphous for the normal structure while polycrystalline for the reversed one, in agreement with the XRD results. Note that the MgO layers in both structures are crystalline, which indicates that the crystallization of the thin top W film in the reversed structure is likely promoted (I) by the bottom MgO layer through a templating effect, which does not take place in the normal structures, and (II) by the fact that the W layer is not adjacent to the $SiO_2$ layer (like in the normal structure), thus preventing any possible interfacial oxide formation. This allows the deposition of α-W films in the thin regime. Nevertheless, polar magneto-optical Kerr effect measurements and anomalous Hall effect (AHE) measurements (both carried out at ambient temperature) show that all the annealed normal and reversed heterostructures possess PMA. Representative out-of-plane hysteresis loops of the W(4)/CoFeB(1.0)/MgO(2) and MgO(2)/CoFeB(1.2)/W(4) samples obtained by AHE measurements are displayed in Figure 1e and f, respectively. The reversed structure shows a sharper hysteresis loop than the normal structure indicating a better crystallinity of both the CoFeB(1.2) layer and the α-W(4) layer in the reversed structure[18].

After investigating their structural and magnetic properties, the samples are patterned into micron-sized Hall-bar devices by photolithography and Ar ion-mill etching (Figure 2a and 2b). We perform current-induced hysteresis loop shift measurement[19] on both series of structures with various W thicknesses to estimate the thickness dependence of the damping-like spin-orbit torque (DL-SOT) efficiency. In this type of measurement, we sweep the out-of-plane magnetic field ($H_z$) while applying a current along the x-direction ($I_{dc}$) and an in-plane bias field ($H_x$) parallel to $I_{dc}$ (Figure 2a and 2b). Within the CoFeB layers, the effective field originated from a Dzyaloshinskii-Moriya interaction (DMI) at the W/CoFeB or CoFeB/W interfaces can be overcome by applying $H_x$ such that the chiral domain wall (DW) magnetic moments will be realigned[9, 10, 20, 21]. When the applied current flows in the W layer, the SHE-induced transverse spin current will generate a SOT acting on the realigned DW moments, inducing an effective out-of-plane field ($H_z^{eff}$) that causes the out-of-plane hysteresis loop to shift along $H_z$, in a direction that depends on the applied current polarity. When $|H_x| \geq |H_{DMI}|$, the shifting effect of the hysteresis loop is maximized and the full strength of the SOT can be detected[19]. Therefore, the saturated DL-SOT efficiency can be estimated by measuring the effective field under different applied currents with $|H_x| \geq |H_{DMI}|$. Representative hysteresis loop shifts are shown in Figure 2c and d for both sample structures and a linear relation between the

detected $H_z^{eff}$ and the applied current are shown in Figure 2e and f. As expected from the different structure stack symmetries, the SOT from the W layers acting on the adjacent CoFeB magnetic DWs is opposite, leading to opposite slope and loop shift direction.

The saturated DL-SOT efficiency of these W-based heterostructures can be estimated by[19]

$$\xi_{DL} = \frac{2e}{\hbar}\left(\frac{2}{\pi}\right)\mu_0 M_s t_{CoFeB}\left(\frac{H_z^{eff}}{J_e}\right), \quad (1)$$

where $M_s$ is the saturation magnetization of the CoFeB layer (characterized by vibrating sample magnetometer (VSM) to be $M_s \approx 1220$ emu/cm$^3$) and $J_e$ is the current density flowing through the spin Hall material (W). However, the current density here cannot be directly calculated by $I_{dc}/(wt_W)$, with $w = 5$ μm being the Hall-bar width, since the applied current not only flows within the W layer but also within the CoFeB and the other buffer layers (in the reversed structure). Therefore, a more accurate current density is calculated by $J_e = J_{dc}\left[\rho_0 t_W/(\rho_0 t_W + \rho_W t_0)\right]$, where $J_{dc} = I_{dc}/(wt_W)$. $t_0$ and $\rho_0$ represent the thickness and resistivity of the overall layers except for the W and the oxidized capping layers. Note that the current only flows within the W and CoFeB layers in the normal structures, which means that $t_0 = t_{CoFeB}$ and $\rho_0 = \rho_{CoFeB} \approx 200$ μΩ-cm. On the other hand, to estimate $\rho_0$ in the reversed structures, we measure the resistance of a sample without a W layer, namely Ta(5)/CoFeB(0.3)/MgO(2)/CoFeB(1.2)/Ta(3), and obtain $\rho_0 \approx 130$ μΩ-cm. Hence, the formula to calculate the DL-SOT efficiency can be rewritten as

$$\xi_{DL} = \frac{2e}{\hbar}\left(\frac{2}{\pi}\right)\mu_0 M_s t_{\text{CoFeB}} w t_{\text{W}} \left(\frac{\rho_0 t_{\text{W}} + \rho_{\text{W}} t_0}{\rho_0 t_{\text{W}}}\right)\left(\frac{H_z^{\text{eff}}}{I_{dc}}\right). \tag{2}$$

By using the values mentioned above, the estimated DL-SOT efficiencies of both structures are summarized in **Figure 3**a where a clear W-phase-dependent trend can be observed. For the normal structure, $t_w \leq 4$ nm is mainly dominated by amorphous W and the DL-SOT efficiency can reach as high as $|\xi_{DL}| \approx 0.20$, whereas within the $5\text{ nm} \leq t_w \leq 6\text{ nm}$ range the W layer experiences a transition from amorphous W to crystalline α-W with intermediate $|\xi_{DL}| \approx 0.10$, while, finally, for $t_w \geq 7$ nm the layer shows a pure α-W phase with $|\xi_{DL}| \approx 0.04$. However, for the reversed structure, only the $t_w = 2$ nm sample shows a purely amorphous W phase with $|\xi_{DL}| \approx 0.14$, whereas all the $t_w \geq 3$ nm samples consist of polycrystalline α-W with $|\xi_{DL}| \leq 0.04$. These results suggest that very thin α-W layers with decent crystallinity can be readily achieved by depositing them in the reversed order, and the SOT-driven switching of an adjacent FM layer should be feasible due to the non-negligible $\xi_{DL}$.

We further characterize the spin Hall conductivity of W by analyzing the relation between $|\xi_{DL}|$ and the measured W layers resistivity ( $\rho_W$ ). The spin Hall conductivity $|\sigma_{SH}| = |\xi_{DL}|/\rho_W$ can be extracted from the linear slope of $|\xi_{DL}|$ vs. $\rho_W$ plot if the observed SHE is intrinsic in nature[22-24]. Interestingly, there exists two linear regimes, which correspond to $|\sigma_{SH}^{\alpha\text{-W}}| \approx 3.71 \times 10^5$ $\Omega^{-1}\text{m}^{-1}$ for α-W-dominated samples and $|\sigma_{SH}^{\text{amorphous-W}}| \approx 1.05 \times 10^5$ $\Omega^{-1}\text{m}^{-1}$ for amorphous-W-dominated samples (Figure 3b). The spin Hall conductivity extracted from our α-W films is of the same order but greater in magnitude compared to a previous first-principles study wherein $\sigma_{SH}^{W} \approx -1.4 \times 10^5$ $\Omega^{-1}\text{m}^{-1}$ was predicted for crystalline W[25]. The fact that α-W has a higher spin Hall conductivity than amorphous W suggests that it is possible to further increase $|\xi_{DL}|$ to > 0.20 by tuning the resistivity of α-W. Although beyond the scope of the present study, possible approaches include doping of nitrogen[26] or oxygen[12, 27] into α-W, or tuning the gas pressure during sputter deposition[28]. Also note that the $|\sigma_{SH}^{\alpha\text{-W}}|$ value reported here is already comparable to that of Pt, $\sigma_{SH}^{Pt} \sim 10^5 \ \Omega^{-1}\text{m}^{-1}$ [22, 23, 29].

Furthermore, we demonstrate current-induced SOT-driven magnetization switching in both normal and reversed film structures. To achieve SOT-driven switching, we apply current pulses with $0.01 \text{ s} \leq t_{\text{pulse}} \leq 1 \text{ s}$ to generate the SHE and an external in-plane field ($H_x$) to overcome the DMI effective field. The current-induced magnetization switching results of the W(4) devices for both the normal and reversed structures are shown in Figure 3c and d, respectively. Note that W(4) is amorphous in the normal structure and crystalline (α-W) in the reversed case. It has been shown in previous studies that amorphous or β-W-based normal W/CoFeB/MgO structures can be switched by currents,[11, 18] while it has not been reported for α-W-based heterostructures yet. Figure 3d shows that current-induced switching can be realized utilizing the SHE from a α-W layer in a reversed structure. Since, in the latter case, α-W can be deposited in the thin regime, the current density flowing in α-W films in the reversed structure is higher than that in thicker α-W films deposited in the normal structure. In general, the higher the current density, the smaller the critical switching current and, therefore, successful magnetization switching could be achieved before the Hall-bar device is destroyed by applying large currents. Note, in Figure 3c and d, that the opposite switching polarity of the CoFeB films results from the different deposition order of the two heterostructures' layers.

Moreover, since current-induced magnetization switching is a thermally-activated process at these current pulse lengths, the switching current $I_c$ depends on the applied pulse width, and can be expressed as[30]

$$I_c = I_{c0}\left[1 - \frac{1}{\Delta}\ln\left(\frac{t_{\text{pulse}}}{\tau_0}\right)\right], \qquad (3)$$

where $I_{c0}$ is the zero-thermal critical switching current, $\Delta \equiv U/k_B T$ is the thermal stability factor ($U$ is the energy barrier), and $1/\tau_0$ ($\tau_0 \approx 1\,\text{ns}$) is the intrinsic attempt frequency[31]. The results of the W(4) devices for both structures are shown in Figure 3e and f. As displayed, linear trends can be observed in the $I_c$ vs $\ln(t_{\text{pulse}}/\tau_0)$ plot. By performing linear fits of the switching data, we estimate $\Delta \approx 36.2$ with $|J_{c0}| \approx 1.54 \times 10^{11}$ A/m² for the normal structure device and $\Delta \approx 25.4$ with $|J_{c0}| \approx 6.63 \times 10^{11}$ A/m² for the reversed structure device. The lower thermal stability factor indicates that MgO/CoFeB/α-W has a lower retention while compared to the normal W/CoFeB/MgO structure, however further layer stack optimization should improve this property.

    To summarize, we verify through XRD measurements and TEM imaging that low-resistive and crystalline α-W in the reversed structure MgO/CoFeB/W can be achieved in thin regime ($t_w \approx 4$ nm) due to a templating effect from its buffer layers. Through current-induced hysteresis loop shift measurements, we show that the estimated DL-SOT efficiency strongly depends on the structural phase of W layer. Moreover, the phase-dependent spin Hall conductivity of W is found to have two distinct regimes: $|\sigma_{\text{SH}}^{\alpha\text{-W}}| \approx 3.71 \times 10^5$ $\Omega^{-1}\text{m}^{-1}$ and $|\sigma_{\text{SH}}^{\text{amorphous-W}}| \approx 1.05 \times 10^5$ $\Omega^{-1}\text{m}^{-1}$. We further demonstrate current-induced DL-SOT switching using normal-structure amorphous-W/CoFeB/MgO devices and, for the first time, in reversed-structure MgO/CoFeB/α-W devices. Our discovery of low-resistive and polycrystalline α-W thin films possessing high spin Hall conductivity and enabling current-induced DL-SOT switching of an adjacent ferromagnet suggests that this HM material could be used as an efficient SOT source for future spintronic applications with suitable thin film engineering.

**Experimental Section**

The film stacks were deposited onto thermally-oxidized Si(001) substrates using a ultra-high vacuum custom-built magnetron sputtering chamber with a base pressure of ~ $1\times10^{-9}$ Torr. All the films were deposited at ambient temperature in an Ar pressure of 3 mTorr using either DC or RF magnetron sputtering, except for the Ta layers that were deposited using ion beam sputtering using Kr gas. The film compositions were determined using Rutherford backscattering spectrometry. Thermal post-annealing of all the films was carried out using an anneal furnace with a base pressure of ~ $2\times10^{-8}$ Torr at 300 °C for one hour with a 1 T magnetic field applied in the direction perpendicular to the films' surface. Magnetic properties were measured by a homebuilt anomalous Hall voltage probe station and the magnetizations were further characterized by a vibrating sample magnetometer (VSM). XRD $\theta$-$2\theta$ out-of-plane measurements were performed at ambient temperature using a Bruker general area detector diffraction system (GADDS) system. TEM images were obtained by a JOEL 2010F field-emission transmission electron microscope (FE-TEM). Resistivities of all films were characterized by four-point measurements. Micron-sized Hall-bar samples for spin-orbit torque characterization and current-induced magnetization switching measurements were prepared by standard photolithography, followed by an Ar ion-mill patterning process.

**Acknowledgements**

This work is supported by the Ministry of Science and Technology of Taiwan (MOST) under grant No. 105-2112-M-002-007-MY3 and 108-2636-M-002-010- and by the Center of Atomic Initiative for New Materials (AI-Mat), National Taiwan University from the Featured Areas Research Center Program within the framework of the Higher Education Sprout Project by the Ministry of Education (MOE) in Taiwan under grant No. NTU-107L9008. We thank Yi-Da Wang for his assistance on TEM imaging, Tsao-Chi Chuang for her assistance on VSM measurements and Holt Bui for the RBS measurements carried out at IBM Almaden Research Center.

**Conflict of Interest**

**References**

[1]     M. I. Dyakonov, V. I. Perel, Phys. Lett. A **1971**, 35, 459.
[2]     J. E. Hirsch, Phys. Rev. Lett. **1999**, 83, 1834.
[3]     A. Hoffmann, IEEE Trans. Magn. **2013**, 49, 5172.
[4]     J. Sinova, S. O. Valenzuela, J. Wunderlich, C. H. Back, T. Jungwirth, Rev. Mod. Phys. **2015**, 87, 1213.
[5]     I. M. Miron, K. Garello, G. Gaudin, P. J. Zermatten, M. V., Costache, Stéphane Auffret, Sebastien Bandiera, Bernard Rodmacq, Alain Schuhl, P. Gambardella, Nature **2011**, 476, 189.
[6]     L. Q. Liu, C.-F. Pai, Y. Li, H. W. Tseng, D. C. Ralph, R. A. Buhrman, Science **2012**, 336, 555.
[7]     L. Q. Liu, C.-F. Pai, D. C. Ralph, R. A. Buhrman, Phys. Rev. Lett. **2012**, 109, 186602.
[8]     V. E. Demidov, S. Urazhdin, H. Ulrichs, V. Tiberkevich, A. Slavin, D. Baither, G. Schmitz, S. O. Demokritov, Nat. Mater. **2012**, 11, 1028.
[9]     S. Emori, U. Bauer, S. M. Ahn, E. Martinez, G. S. D. Beach, Nat. Mater. **2013**, 12, 611.
[10]    K. S. Ryu, L. Thomas, S. H. Yang, S. Parkin, Nat. Nanotechnol. **2013**, 8, 527.
[11]    C.-F. Pai, L. Q. Liu, Y. Li, H. W. Tseng, D. C. Ralph, R. A. Buhrman, Appl. Phys. Lett. **2012**, 101, 122404.
[12]    K. U. Demasius, T. Phung, W. F. Zhang, B. P. Hughes, S. H. Yang, A. Kellock, W. Han, A. Pushp, S. S. P. Parkin, Nat. Commun. **2016**, 7, 10644.
[13]    Q. Hao, W. Chen, G. Xiao, Appl. Phys. Lett. **2015**, 106, 182403.
[14]    J. Liu, T. Ohkubo, S. Mitani, K. Hono, M. Hayashi, Appl. Phys. Lett. **2015**, 107, 232408.
[15]    L. Zhu, D. C. Ralph, R. A. Buhrman, Phys. Rev. Applied **2018**, 10, 031001.
[16]    M.-H. Nguyen, S. Shi, G. E. Rowlands, S. V. Aradhya, C. L. Jermain, D. C. Ralph, R. A. Buhrman,  **2018**, 112, 062404.
[17]    N. H. D. Khang, Y. Ueda, P. N. Hai, Nat. Mater. **2018**, 17, 808.
[18]    T. C. Wang, T. Y. Chen, H. W. Yen, C. F. Pai, Phy Rev Mater **2018**, 2, 014403.
[19]    C. F. Pai, M. Mann, A. J. Tan, G. S. D. Beach, Phys. Rev. B **2016**, 93, 144409.
[20]    O. J. Lee, L. Q. Liu, C. F. Pai, Y. Li, H. W. Tseng, P. G. Gowtham, J. P. Park, D. C. Ralph, R. A. Buhrman, Phys. Rev. B **2014**, 89, 024418.
[21]    S. Emori, E. Martinez, K. J. Lee, H. W. Lee, U. Bauer, S. M. Ahn, P. Agrawal, D. C. Bono, G. S. D. Beach, Phys. Rev. B **2014**, 90, 184427.
[22]    M. Morota, Y. Niimi, K. Ohnishi, D. H. Wei, T. Tanaka, H. Kontani, T. Kimura, Y. Otani, Phys. Rev. B **2011**, 83, 174405.
[23]    E. Sagasta, Y. Omori, M. Isasa, M. Gradhand, L. E. Hueso, Y. Niimi, Y. Otani, F. Casanova, Phys. Rev. B **2016**, 94, 060412.

[24] E. Sagasta, Y. Omori, S. Vélez, R. Llopis, C. Tollan, A. Chuvilin, L. E. Hueso, M. Gradhand, Y. Otani, F. Casanova, Phys. Rev. B **2018**, 98, 060410.
[25] Y. Yao, Z. Fang, Phys. Rev. Lett. **2005**, 95, 156601.
[26] T. Y. Chen, C. T. Wu, H. W. Yen, C. F. Pai, Phys. Rev. B **2017**, 96, 104434
[27] H. An, Y. Kanno, A. Asami, K. Ando, Phys. Rev. B **2018**, 98, 014401.
[28] J. W. Lee, Y.-W. Oh, S.-Y. Park, A. I. Figueroa, G. van der Laan, G. Go, K.-J. Lee, B.-G. Park, Phys. Rev. B **2017**, 96, 064405.
[29] M. H. Nguyen, D. C. Ralph, R. A. Buhrman, Phys. Rev. Lett. **2016**, 116, 126601.
[30] R. H. Koch, J. A. Katine, J. Z. Sun, Phys. Rev. Lett. **2004**, 92, 088302.
[31] A. V. Khvalkovskiy, D. Apalkov, S. Watts, R. Chepulskii, R. S. Beach, A. Ong, X. Tang, A. Driskill-Smith, W. H. Butler, P. B. Visscher, D. Lottis, E. Chen, V. Nikitin, M. Krounbi, J. Phys. D: Appl. Phys. **2013**, 46, 074001.

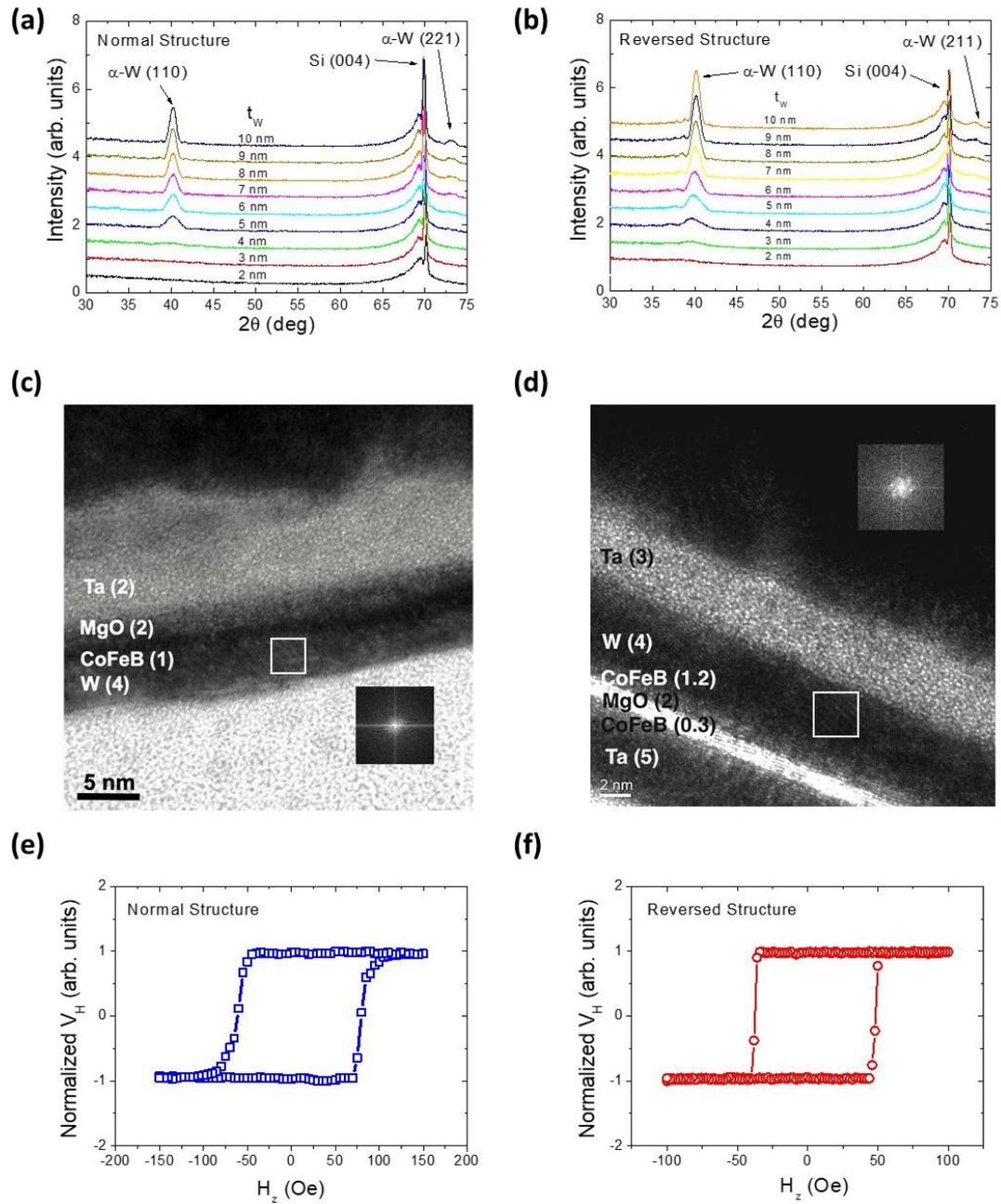

**Figure 1**. θ-2θ XRD data of a) the normal W($t_w$)/CoFeB(1)/MgO(2) and b) reversed Ta(5)/CoFeB(0.3)/MgO(2)/CoFeB(1.2)/W($t_w$) structures. Cross-sectional TEM images of the c) W(4)/CoFeB(1)/MgO(2) and d) Ta(5)/CoFeB(0.3)/MgO(2)/CoFeB(1.2)/W(4) magnetic heterostructures. The subpanels represent the diffractograms obtained from the FFT analysis of the W(4) layers. Representative AHE hysteresis loops of the e) W(4)/CoFeB(1)/MgO(2) and f) Ta(5)/CoFeB(0.3)/MgO(2)/CoFeB(1.2)/W(4) samples measured at room temperature.

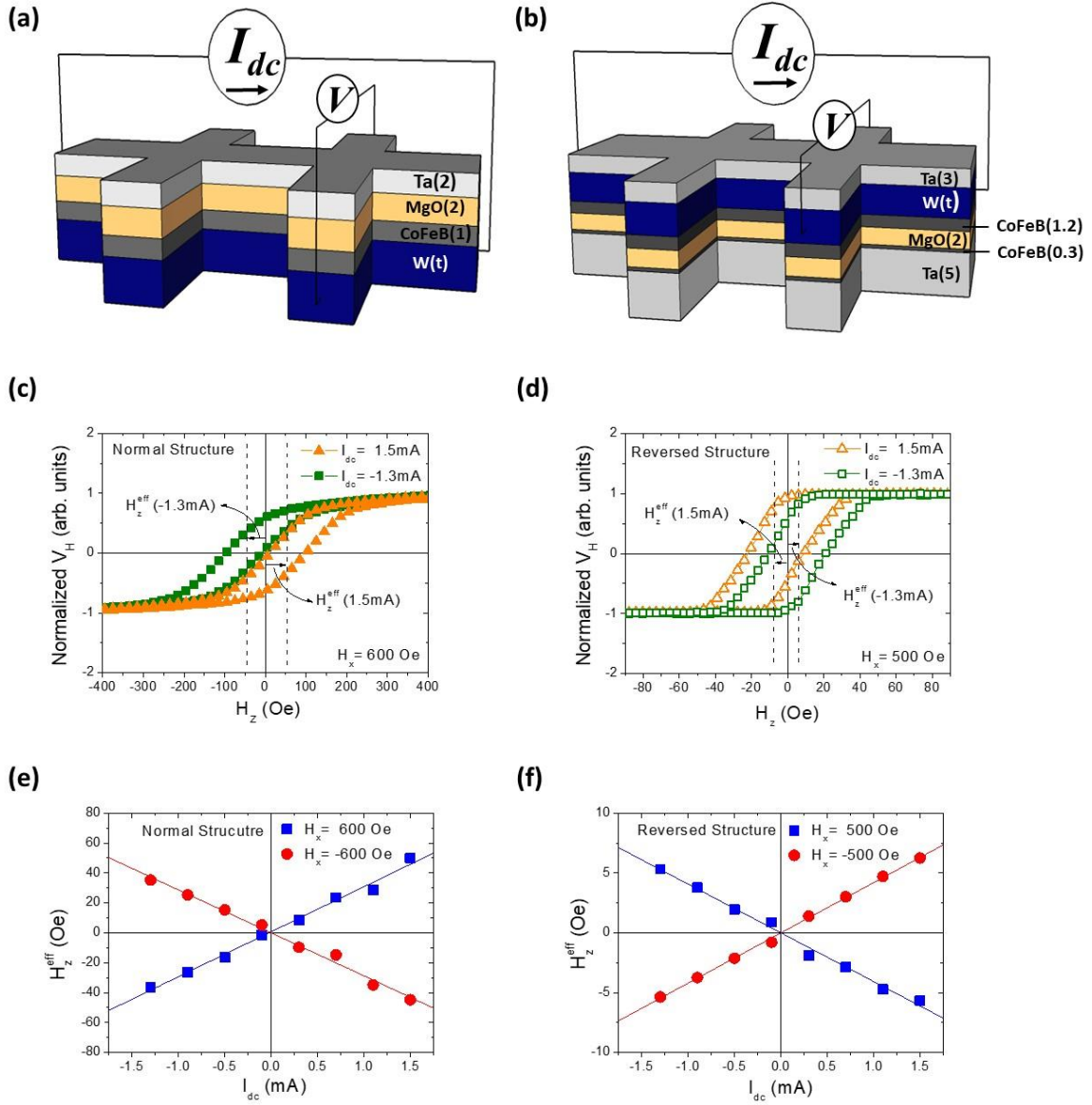

**Figure 2**. Schematic illustration of the anomalous Hall voltage measurements of a) the normal and b) the reversed structures. Representative shifted out-of-plane hysteresis loops of the c) W(4)/CoFeB(1)/MgO(2) and d) Ta(5)/CoFeB(0.3)/MgO(2)/CoFeB(1.2)/W(4) devices. Current-induced effective field $H_z^{\text{eff}}$ of the e) W(4)/CoFeB(1)/MgO(2) device under $H_x = \pm 600$ Oe and the f) Ta(5)/CoFeB(0.3)/MgO(2)/CoFeB(1.2)/W(4) device under $H_x = \pm 500$ Oe.

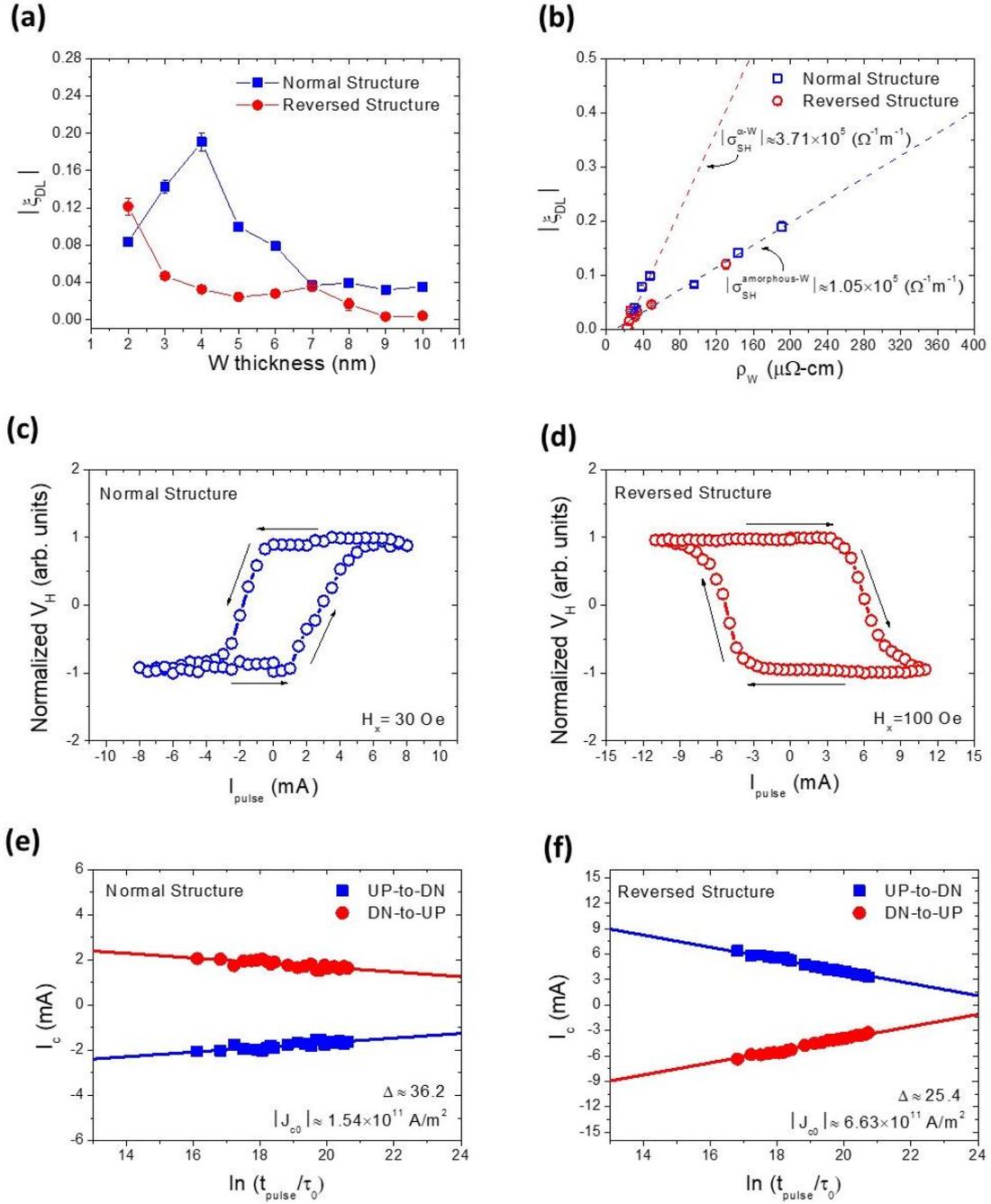

**Figure 3.** The magnitude of the DL-SOT efficiency $|\xi_{DL}|$ vs a) W thickness and b) resistivity of the normal (blue squares) and the reversed (red circles) structures. The dash lines in (b) represent linear fits to the data. Representative current-induced switching results of the c) W(4)/CoFeB(1)/MgO(2) and d) Ta(5)/CoFeB(0.3)/MgO(2)/CoFeB(1.2)/W(4) devices. The black arrows represent the sweeping directions of the applied current pulses. The applied current pulse width dependence of the switching currents for the e) W(4)/CoFeB(1)/MgO(2) and f) Ta(5)/CoFeB(0.3)/MgO(2)/CoFeB(1.2)/W(4) devices. The solid lines represent linear fits to the experimental data. DN-to-UP and UP-to-DN represent magnetization switching from $-z$ to $+z$ and $+z$ to $-z$, respectively.